\documentclass[prd,twocolumn,showpacs,amsmath,amssymb]{revtex4}

\usepackage{amssymb}
\usepackage{mathrsfs}
\usepackage{txfonts}

\usepackage{graphicx}
\usepackage{dcolumn}% Align table columns on decimal point
\usepackage{bm}% bold math

\begin{document}

\title{Gravitational wave from warm inflation}

\author{Xi-Bin Li}
\email{lxbbnu@mail.bnu.edu.cn}
\affiliation{Department of Physics, Beijing Normal University, Beijing 100875, China}

\author{He Wang}
\email{hewang@mail.bnu.edu.cn}
\affiliation{Department of Physics, Beijing Normal University, Beijing 100875, China}

\author{Jian-Yang Zhu}
\thanks{Corresponding author}
\email{zhujy@bnu.edu.cn}
\affiliation{Department of Physics, Beijing Normal University, Beijing 100875, China}

\date{\today}

\begin{abstract}
A fundamental prediction of inflation is a nearly scale-invariant spectrum of gravitational wave. The features of such a signal provide extremely important information about the physics of the early universe. In this paper, we focus on several topics about warm inflation. First, we discuss the stability property about warm inflation based on nonequilibrium statistical mechanics, which gives more fundamental physical illustrations to thermal property of such model. Then, we calculate the power spectrum of gravitational waves generated during warm inflation, in which there are three components contributing to such spectrum: thermal term, quantum term and cross term combining the both. We also discuss some interesting properties about these terms and illustrate them in different panels. As a model different from cold inflation, warm inflation model has its individual properties in observational practice, so we finally give a discussion about the observational effect to distinguish it from cold inflation.
\end{abstract}


\maketitle

%\tableofcontents

\section{\label{introduction}Introduction}

It has taken for several years to search for gravitational waves on astrophysical experiments, such as LIGO \cite{PhysRevX.6.041015}, VIRGO \cite{VIRGO} and other observational experiments \cite{GEO600,TAMA300} to test the prediction of general relativity. Hulse and Taylor \cite{1975ApJ...195L..51H} first announced the observational evidence for the existence of gravitational waves. Then came the good news that gravitational waves have been detected by LIGO \cite{Abbott2016, PhysRevLett.116.061102, PhysRevLett.116.241103} in recent years generated by binary black hole. This discovery was a great achievement that opened a new window to better understand our universe both at early epoch and late epoch.

Warm inflation model was established as a candidate scenario to overcome some defects in cold inflation \cite{doi:10.1142/S0217751X09044206,BARTRUM2014116}. However, it was realized a few years after its original proposal that the idea of warm inflation was not easy to realize in concrete models and even simply not be possible in relevant work \cite{PhysRevD.60.083509,PhysRevLett.83.264}. Some problems were mentioned to suspect such scenario. First, it is hard to couple the inflaton directly with light fields. Considering a Yukawa interaction $g\phi\bar{\psi}\psi$, the slow roll condition typically requires an inflaton with large value, while the fermion obtain an extra mass with also a large value unless the coupling a quite small. As a result, the dissipative effect may be too small either which implies that it is hard to obtain a period of inflation with dissipation coefficient long enough to require the 60 e-ford to solve the horizon and flatness problem. Second, a direct coupling to light fields may lead to large thermal corrections to the inflaton mass $m_\psi=g\phi$, which could prevent slow roll for $T>H$. Shortly afterwards successful models of warm inflation have been established, in which the inflaton indirectly interacts with the light degrees of freedom though a heavy mediator fields instead of being coupled with a light field directly \cite{doi:10.1142/S0217751X09044206,PhysRevD.84.103503,PhysRevLett.117.151301,1475-7516-2011-09-033}. In such scenarios dissipation can sustain both the slow-roll dynamics of the inflaton field and the temperature of the radiation bath for a sufficiently long period. One can read a Lagrangian density of the generic form,
\begin{eqnarray}
    \mathcal{L}[\Phi,X,Y]=\mathcal{L}[\Phi]& &+\mathcal{L}[X]+\mathcal{L}[Y]\nonumber \\& &+\mathcal{L}_\textrm{int}[\Phi,X]+\mathcal{L}_\textrm{int}[X,Y]
\end{eqnarray}
where $\Phi$ is the inflaton field, $X$ are any field or degrees of freedom coupled directly to the inflaton field, while $Y$ can be any other fields not necessarily coupled to the inflaton, but are coupled to $X$. The $\mathcal{L}_\textrm{int}[\Phi,X]$ and $\mathcal{L}_\textrm{int}[X,Y]$ give the interaction among these fields. The evolution of the inflaton field can be properly determined in the context of the in-in, or the Schwinger closed-time path functional formalism \cite{978-0-521-87499-1}. This equation displays both dissipation and non-Markovian stochastic noise terms and it is a generalized Langevin-like equation of motion \cite{PhysRevD.76.083520,PhysRevD.91.083540}.

Compared with the predictions of cold inflation that primordial density fluctuations mostly from quantum fluctuation and thermal bath are only generated at the end of inflation \cite{RevModPhys.78.537}, warm inflation model suggests that our universe is hot during the whole inflation when inflaton fields couple with the thermal bath and the primary source of density fluctuations come from thermal fluctuations \cite{PhysRevD.64.063513, PhysRevD.62.083517, PhysRevD.69.083525}. The equation of motion for warm inflation can be written as a stochastic Lengevin equation, in which there is a dissipation term to describe the inflaton fields coupling with thermal bath and there is also a fluctuation term described by a stochastic noise term \cite{PhysRevD.71.023513, PhysRevD.76.083520}. The fundamental principles of warm inflation have been reviewed recently in \cite{PhysRevD.50.2441}. Warm inflation with strong dissipation requires $\Upsilon>H$, with which thermal fluctuations dominate the whole inflationary epoch i.e. $T>H$ \cite{PhysRevD.58.123508}. Generally, we can still have the weak dissipation condition with $\Upsilon\ll H$ \cite{PhysRevD.59.083503, PhysRevD.57.741}. Noncanonical warm inflation model is under studying recently \cite{Zhang1, Zhang2}.

Although primordial gravitational waves generated during inflation have not been detected yet till now, the discovery of gravitational waves by LIGO has shed a bright light to this prediction. The measurement of the cosmic microwave background radiation and other observations have given a good constraint to the ratio of tensor to scaler with $r<0.07$ at $95$ CL \cite{Planck}. The study of primordial gravitational wave is a way to prove the inflationary programm. What's more, it also provides an effective method to distinguish among different inflationary model. Now more and more researchers have cast their eyes to primordial gravitational waves generated by various kind of sources, such as primordial density perturbation \cite{PhysRevD.76.084019, PhysRevD.75.123518}, some kind of inflationary model \cite{PhysRevD.85.023534, PhysRevD.88.103518}, and even during reheating epoch \cite{1475-7516-2016-02-023}; a great deal of predictions have been given in these works which may be observed in observation experiments in the near future. However, till now there has not been any calculation about the gravitational waves generated from warm inflation. This approach must be an effective way to differentiate cold inflation model from warm inflation model.

In this paper, we attempt to illustrate the existence of gravitational wave generated from warm inflation. Green's function method has been used to calculate the power spectrum of tensor perturbation with source as the form of transverse-traceless tensor. This spectrum can separate
into three terms: thermal component, quantum component and cross component, and each component has its individual and interesting properties. With these properties, we discuss the observational prediction from the view of tensor-to-scalar ratio.

This paper is organized as follows: in Sec. \ref{Stochastic_approach}, we give a brief introduction to warm inflation and stochastic approach to  deal with warm inflation. In Sec. \ref{spectrum}, based on nonequilibrium statistical mechanics, we recalculate the statistical properties of warm inflation model and give more results in detail. In Sec. \ref{GW} three programs for describing the thermal, quantum and their cross term are calculated and relevant calculations and equations are derived. In Sec. \ref{result}, we discuss our result by numerical analysis. Finally, in Sec. \ref{conclusion}, we conclude our work and give some further discussions about our results.

\section{\label{Stochastic_approach}Stochastic approach for warm inflation}
\subsection{\label{StochasticApproach}Stochastic approach}

First, let's have a brief review of stochastic approach for cold inflation. The equation of motion for cold inflation is the standard one:
In warm inflation model, the equation of background field is often written as the Langevin equation
\begin{eqnarray}
\Big[\frac{\partial^2 }{\partial t^2}+3H\frac{\partial}{\partial t}-\frac{1}{a^2}\nabla^2\Big]\Phi+\frac{\partial V(\Phi)}{\partial \Phi} = 0, \label{EOMcoldinflaton}
\end{eqnarray}
where $\Phi$ is the inflaton field operator, $a$ is scale factor in Friedmann-Robertson-Walker metric and $H$ is the Hubble parameter defended as $H=\dot{a}/a$. The stochastic approach assumes that inflaton field separates into two parts, one is $\Phi_{>}$ which denotes the long wavelength part, and another is $\Phi_{<}$ which denotes the short wavelength part for quantum vacuum fluctuation, i.e. $\Phi\rightarrow\Phi_{>}+\Phi_{<}$. Usually, $\Phi_<$ is written in terms of a filter (window) function.

The stochastic inflationary approach suggests that quantum inflaton field $\Phi$ is composed in a short wavelength part $\Phi_<$, which denotes the quantum vacuum fluctuations, and a long wavelength part $\Phi_>$ i.e. $\Phi(\textbf{x},t)=\Phi_<(\textbf{x},t)+\Phi_>(\textbf{x},t)$. As previously study, quantum fluctuation dominates on short wavelength. With this condition, we assume a number smaller than unity such that the quantum
\begin{eqnarray}
\Phi_{<}(\textbf{x},t)\equiv\phi_q(\textbf{x},t)=& & \int{\frac{d^3k}{(2\pi)^{3/2}}W(k,t)}   \nonumber\\
& & \times\Big[\phi_{\textbf{k}}(t)e^{-i\textbf{x}\cdot\textbf{k}}\hat{a}_\textbf{k}+\textrm{h.c.}\Big], \label{quanforshort}
\end{eqnarray}
where $\phi_{\textbf{k}}(t)$ is the field in momentum space, $\hat{a}_\textbf{k}$ is the annihilation operator whose Hermitian conjugate operators is $\hat{a}^\dag_\textbf{k}$. In ~(\ref{quanforshort}), $W(k,t)$ are the window function with sharp momentum cutoff
\begin{eqnarray}
W(k,t)=\theta(k-\epsilon aH). \label{windowfunction}
\end{eqnarray}
where $\epsilon$ is a suitable number smaller than 1. Then fields $\phi_{\textbf{k}}(t)$ in \eqref{quanforshort} satisfy
\begin{eqnarray}
\Big[\frac{\partial^2 }{\partial t^2}+3H\frac{\partial}{\partial t}-\frac{k^2}{a^2}+\langle V,_{\phi\phi}(\Phi_{>})\rangle\Big]\phi_{\textbf{k}}(t) = 0. \label{quantumvacuum}
\end{eqnarray}
Generally, $\phi_{\textbf{k}}(\tau)$ is given by
\begin{eqnarray}
\phi_{\textbf{k}}(\tau)=\frac{H\sqrt{\pi}}{2}(-\tau)^{3/2}H_\mu^{(1)}(-k\tau). \label{solution_phi_k}
\end{eqnarray}
where $\mu=\sqrt{9/4-m^2/H^2}$ with $m^2\simeq$ constant is the average of $V,_{\phi\phi}(\Phi_{>})$, $\tau$ is the conformal time in de Sitter space-time defined as $\tau=-1/aH$, and $H^{(1)}$ is the Hankel function of the first kind.

Then, with ~(\ref{EOMcoldinflaton}), the equation of motion for long wavelength part of the field $\Phi_{>}$ reads
\begin{eqnarray}
     \Big[\frac{\partial^2 }{\partial t^2}+3H\frac{\partial}{\partial t}-\frac{1}{a^2}\nabla^2\Big]\Phi_{>}+{V,_{\phi}(\Phi_{>})} = \xi_q, \label{EOMlongwave}
\end{eqnarray}
where $\xi_q$ is given by
\begin{eqnarray}
     \xi_q=-\Big[\frac{\partial^2 }{\partial t^2}+3H\frac{\partial}{\partial t}-\frac{1}{a^2}\nabla^2+ V,_{\phi\phi}(\Phi_{>})\Big]\phi_{\textbf{q}}(t) \label{xiq}
\end{eqnarray}
and its average vanishes $\langle\xi_q\rangle=0$. With ~(\ref{quanforshort}) and ~(\ref{EOMlongwave}), the two-point correlation function is
\begin{eqnarray}
     \langle\xi_q(\textbf{x},t)\xi_q(\textbf{x}',t')\rangle = \int{\frac{d^3k}{(2\pi)^3}e^{i\textbf{k}\cdot(\textbf{x}-\textbf{x}')}\textrm{Re}\Big[f_\textbf{k}(t)f_{\textbf{k}'}^*(t')\Big]}, \nonumber\\ \label{corelationfunction}
\end{eqnarray}
where
\begin{eqnarray}
     f_\textbf{k}(t)=\Big[\frac{\partial^2 W}{\partial t^2}+3H\frac{\partial W}{\partial t}\Big]\phi_\textbf{k}(t)+2 \frac{\partial W}{\partial t}\frac{\partial \phi_\textbf{k}(t)}{\partial t}. \label{fk}
\end{eqnarray}
Here we just give a brief introduction of stochastic inflationary approach, and more details have been studied in several work
\cite{2004JCAP...08..011L, PhysRevD.61.084008}.

\subsection{\label{warm_inflation}Warm inflation}
In warm inflation model, the equation of motion of background field is often written as the Langevin equation
 \begin{eqnarray}
    \Big[\frac{\partial^2 }{\partial t^2}+(3H+\Upsilon)\frac{\partial}{\partial t}-\frac{1}{a^2}\nabla^2\Big]\Phi+\frac{\partial V(\Phi)}{\partial \Phi} = \xi_T, \label{EOMwarminflaton}
\end{eqnarray}
where $\Upsilon$ is the dissipation coefficient and $\xi_T$ is the thermal noise fluctuation. In this paper, we consider only in the case of de Sitter space-time, where $a(t)=\textrm{exp}(Ht)$ and $H$ is regarded as a constant. According to the fluctuation-dissipation theorem, dissipation coefficient $\Upsilon$ and fluctuation noise $\xi_T$ have the relation
 \begin{eqnarray}
    \langle \xi_T(\textbf{x},t)\xi_T(\textbf{x},t')\rangle=2\Upsilon T a^{-3}\delta(t-t'). \label{dis_flu_relation}
\end{eqnarray}
The Fourier transformation of ~(\ref{dis_flu_relation}) is
\begin{eqnarray}
    \langle \xi_T(\textbf{k},t)\xi_T(\textbf{k}',t')\rangle=2(2\pi^3)\Upsilon T a^{-3}\delta^3(\textbf{k}+\textbf{k}')\delta(t-t'). \nonumber\\ \label{dis_flu_relation_k}
\end{eqnarray}
Usually $\Upsilon$ is a function of both background homogeneous inflaton field $\Phi$ and temperature $T$ \cite{1475-7516-2011-09-033} and we do not attend to discuss this question here.

The the inflaton field operator $\Phi(\textbf{x},t)$ is often separated into the parts as follow
\begin{eqnarray}
   \Phi(\textbf{x},t)=\phi(t)+\delta\varphi(\textbf{x},t), \label{inflaton}
\end{eqnarray}
where $\delta\varphi(\textbf{x},t)$ is the perturbed part of inflaton, and $\phi(t)$ is the background homogeneous inflaton field which is defended as
\begin{eqnarray}
   \phi(t)=\frac{1}{\Omega}\int_\Omega{d^3x \Phi(\textbf{x},t)}. \label{BGinflaton}
\end{eqnarray}
Here, $\Omega$ is particle horizon size $\Omega=1/H$. With this relation, ~(\ref{EOMwarminflaton}) reads
\begin{eqnarray}
     \frac{\partial^2 \phi}{\partial t^2}+[3H+\Upsilon]\frac{\partial \phi}{\partial t}+V,_\phi(\phi) &=& 0,\label{EOM_BGinflaton} \\
    \Big\{\frac{\partial^2}{\partial t^2}+[3H+\Upsilon(\phi)]\frac{\partial}{\partial t}-\frac{1}{a^2}\nabla^2+\nonumber \\
    \Upsilon_\phi(\phi)\dot{\phi}
    +V_{\phi\phi}(\phi)\Big\}\delta\varphi &=& \xi_T. \label{EOM_pertubed_field}
\end{eqnarray}

With the slow-roll condition, we write ~(\ref{EOM_BGinflaton}) as form of that in cold inflation,
\begin{eqnarray}
    3H(1+Q)\dot{\phi}+V,_\phi(\phi)=0, \label{Approx_WI}
\end{eqnarray}
where $Q$ is the ratio of dissipation coefficient $\Gamma$ and  Hubble parameter $H$ i.e. $Q\equiv\Gamma/3H$. It is also necessary to define some slow-roll parameter for warm inflation,
\begin{eqnarray}
    \varepsilon &=& \frac{1}{16 \pi G}\Big(\frac{V,_\phi}{V}\Big)^2 \ll 1+Q , \label{varepsilon} \\
    \eta &=& \frac{1}{8 \pi G}\frac{V,_{\phi\phi}}{V} \ll 1+Q , \label{eta}
\end{eqnarray}
and
\begin{eqnarray}
    \beta &=& \frac{1}{8 \pi G}\frac{ \Upsilon,_\phi V,_{\phi}}{\Upsilon V} \ll 1+Q . \label{beta}
\end{eqnarray}

Now, looking again ~(\ref{EOM_pertubed_field}), we consider only the fluctuation from thermal noise and neglect that from quantum noise. Obviously, although thermal effect play a significant role in warm inflation, quantum effect may also be non-negligible. Quantum noise dominates  still for short wavelength perturbation, so we can use stochastic inflationary approach to deal the quantum fluctuation in warm inflation. Thus ~(\ref{EOM_pertubed_field}) in momentum space by defining the variable $z=k/aH$ (ranging from 0 to $\infty$) becomes
\begin{widetext}
\begin{eqnarray}
    \delta\varphi''(\textbf{k},z)-\frac{1}{z}(3Q+2)\delta\varphi'(\textbf{k},z)+\bigg[1+\frac{3(\eta-\beta Q/(1+q))}{z^2}\bigg] \delta\varphi(\textbf{k},z)=\frac{1}{z^2H^2}\big[\xi_q(\textbf{k},z)+\xi_T(\textbf{k},z)\big], \label{EOM_pertubed_field_z}
\end{eqnarray}
\end{widetext}
where primes denote the derivatives with respect to the variable $z$ and $\xi_q(\textbf{k},z)$ is the quantum noise term reads
\begin{eqnarray}
    \xi_q(\textbf{k},z)=-\frac{H^2z^2}{k^2}\Big[& &\frac{\partial ^2}{\partial z^2}-\frac{2+3Q}{z}\frac{\partial}{\partial z}+1\nonumber\\
    & &+\frac{\Upsilon,_{\phi}\dot{\phi}}{H^2z^2}+\frac{V,_{\phi\phi}}{H^2z^2} \Big]\hat{\phi}_q(\textbf{k},z), \label{xiqwarm}
\end{eqnarray}
in which
\begin{eqnarray}
   \hat{\phi}_q(\textbf{k},z)=W(k,z)\Big[\phi_{\textbf{k}}(z)\hat{a}_{-\textbf{k}}+\textrm{h.c.}\Big]. \label{xiqwarm1}
\end{eqnarray}
Here $\hat{\phi}_q(\textbf{k},z)$ are still the quantum field modes that satisfy the equation
\begin{eqnarray}
   \Big[\frac{\partial ^2}{\partial z^2}-\frac{2+3Q}{z}\frac{\partial}{\partial z}+1+\frac{V,_{\phi\phi}}{H^2z^2}\Big]\phi_\textbf{k}(z)=0, \label{equationphiz}
\end{eqnarray}
with the solution in terms of $z$:
\begin{eqnarray}
    \phi_\textbf{k}(z) = \frac{H\sqrt{\pi}}{2k^{3/2}}z^{3/2}H_\mu^{(1)}(z), \label{solutionofphiz}
\end{eqnarray}
where $\mu=\sqrt{9/4-V,_{\phi\phi}/H^2}\approx 3/2-3\eta$. Then, we can get the expression of correlation function of quantum noise
\begin{eqnarray}
    \langle\xi_q(\textbf{k},z)\xi_q(\textbf{k}',z')\rangle = \frac{(2\pi)^3z^2z'^2H^4}{k^2k'^2}[2n(k)+1]\nonumber\\
    \delta^3(\textbf{k}+\textbf{k}')\textrm{Re}\Big[f_\textbf{k}(z)f_{\textbf{k}'}^*(z')\Big], \label{correlationfunctionwarm}
    %\Big[f_{\text{k}}(z)f^*_{\text{k}'}(z')\langle\hat{a}_{-\textbf{k}}\hat{a}^\dag_{\textbf{k}'}\rangle\nonumber\\
    %+f^*_{\text{k}}(z)f_{\text{k}'}(z')\langle\hat{a}^\dag_{\textbf{k}'}\hat{a}_{-\textbf{k}}\rangle\Big], \label{solutionofphiz}
\end{eqnarray}
where $z'=k/a(\tau')H$ and $f_\textbf{k}(z)$ is \cite{1475-7516-2013-03-032}
\begin{eqnarray}
    f_\textbf{k}(z)=k^2\Big[W''- & & \frac{3Q+2}{z}W'-\frac{3\beta Q}{(1+Q)z^2}W\Big]\phi_\textbf{k}(z) \nonumber\\
    & & +2k^2W'\phi'_\textbf{k}(z), \label{fkz}
\end{eqnarray}
and $n(k)$ is the distribution function for high-frequency quantum inflaton which satisfies Bose-Einstein distribution, i.e. $n(k)\equiv\langle\hat{a}^\dag_{\textbf{k}'}\hat{a}_{-\textbf{k}}\rangle=1/[\textrm{exp}((k-\mu)/aT)-1]$ where $\mu$ is the chemical potential to eliminate the divergence of several integrals which will be discussed below.

\section{\label{GW}Gravitational Wave generated from Warm Inflation}

Fluctuations (including thermal and quantum) through the energy-momentum tensors $T_{\mu\nu}$ generates tensor perturbation in Friedmann-Robertson-Walker metrics as
\begin{eqnarray}
g_{\mu\nu}=a^2(\tau)[d\tau^2+(\delta_{ij}+h_{ij})dx^idx^j],  \label{metric}
\end{eqnarray}
where dots denote the derivative with respect to conformal time $\tau$ and $H_{ij}$ is the tensor perturbation with transverse-traceless gauge. Tensor perturbations $h_{ij}$ satisfy the equation of motion
\begin{eqnarray}
\ddot{h}_{ij}+2\frac{\dot{a}}{a}\dot{h}_{ij}-\nabla^2 h_{ij} = \frac{2}{M^2_p}\Pi^{\ \ kl}_{ij}T_{kl},  \label{EOMtensor}
\end{eqnarray}
with $M^{-2}_p\equiv 8\pi G$. Define $\tilde{h}_{ij}\equiv a h_{ij}$, then \eqref{EOMtensor} with Fourier transformation becomes
\begin{eqnarray}
\ddot{\tilde{h}}_{ij}+(k^2-\frac{2}{\tau^2})\tilde{h}_{ij} = \frac{2a}{M^2_p}\Pi^{\ \ kl}_{ij}(\textbf{k})T_{kl}(\textbf{k},\tau)\ .
\label{EOMtensork}
\end{eqnarray}
Here, we have used the relation $\tau=-1/aH$ in de Sitter space-time and $\Pi^{\ \ kl}_{ij}$ are the transverse-traceless projectors which follow \cite{gravitationalwaves, Gravitational...Waves..1}:
\begin{eqnarray}
\Pi^{\ \ kl}_{ij}(\textbf{k})p_{k}p_{l}&=&\bigg(p_i -\frac{k_i(\textbf{p}\cdot\textbf{k})}{k^2}\bigg)
\bigg(p_j -\frac{k_j(\textbf{p}\cdot\textbf{k})}{k^2}\bigg) \nonumber \\
&&-\frac{1}{2}(\delta_{ij}-\hat{k}_i\hat{k}_j)\bigg(p^2 -\frac{(\textbf{p}\cdot\textbf{k})^2}{k^2}\bigg),
\label{projector1}
\end{eqnarray}
\begin{eqnarray}
\Pi^{\ \ kl}_{ij}(\textbf{k})\Pi^{\ \ mn}_{ij}(\textbf{k})p_{k}p_{l}p_{m}p_{n}=
\frac{1}{2} \bigg(p^2 -\frac{(\textbf{p}\cdot\textbf{k})^2}{k^2}\bigg)^2,
\label{projector2}
\end{eqnarray}
and
\begin{eqnarray}
\Pi^{\ \ kl}_{ij}(\textbf{k})k_{l}=0, \label{projector3}
\end{eqnarray}
where $\hat{k}_i=k_i/k$. The solution of \eqref{EOMtensork} is
\begin{eqnarray}
\tilde{h}_{ij}(\textbf{k},\tau)=\frac{2}{M_p^2}\int{d\tau'G_k(\tau,\tau')a(\tau')\Pi^{\ \ kl}_{ij}(\textbf{k})T_{kl}(\textbf{k},\tau')}, \nonumber\\ \label{hofGF}
\end{eqnarray}
where $G_k(\tau,\tau')$ is the Green function. Two linear solutions of \eqref{EOMtensork} are %\cite{2003moco.book.....D}
\begin{subequations}\label{solutions}
\begin{eqnarray}
y_1(\tau)=-k\tau \ \textrm{n}^{(1)}_1(-k\tau)=\Big(1+\frac{i}{k\tau}\Big)e^{-ik\tau} \label{solution1}
\end{eqnarray}
and
\begin{eqnarray}
y_2(\tau)=-k\tau \ \textrm{n}^{(2)}_1(-k\tau)=\Big(1-\frac{i}{k\tau}\Big)e^{ik\tau}, \label{solution2}
\end{eqnarray}
\end{subequations}
where $\textrm{n}^{(1)}_1(z)$ and $\textrm{n}^{(2)}_1(z)$ are the spherical Bessel functions of the third kind \cite{1970hmfw.book.....A}.
According to the method in Appendix \ref{Green_function}, Green's functions as a function of $z$ and $z'$ are
\begin{eqnarray}
G_k(z,z')=\frac{1}{kzz'}\bigg[(1+ & & zz')\textrm{sin}(z'-z)+(z-z') \nonumber\\ & & \times\textrm{cos}(z'-z)\bigg]\theta(z'-z),  \label{solutionGFtensor}
\end{eqnarray}

Energy-momentum tensor for warn inflaton field write \cite{2014JCAP...05..004B}
\begin{eqnarray}
T_{ab}=\partial_a \Phi\partial_b \Phi-g_{ab}(\frac{1}{2}\partial_\mu \Phi\partial^\mu \Phi-V). \label{EMtensor}
\end{eqnarray}
The part on the right hand containing $g_{ab}$ is projected away by $\Pi^{\ \ kl}_{ij}$, and $\phi(t)$ is the space average of inflaton field that contributes nothing to the perturbation. So $\delta \varphi$ are the primary part in $T_{ab}$. Thus, the tensor spectrum reads
\begin{eqnarray}
&&\langle h_{ij}({\bf k},\tau )h_{ij}({\bf k}^{\prime },\tau )\rangle
\nonumber \\
&=&\frac 4{(2\pi )^3a(\tau )^2M_p^4}\int {dz^{\prime }a(\tau ^{\prime
})G_k(\tau ,\tau ^{\prime })}\int {dz^{\prime \prime }a(\tau ^{\prime \prime
})G_k(\tau ,\tau ^{\prime \prime })}  \nonumber \\
&&\times \int {d^3pd^3p^{\prime }}\Pi _{ij}^{\ \ kl}({\bf k})\Pi _{ij}^{\ \
mn}({\bf k}^{\prime })p_k(k_l-p_l)p_m^{\prime }(k_n^{\prime }-p_n^{\prime })
\nonumber \\
&&\times \langle \Phi ({\bf p},\tau ^{\prime })\Phi ({\bf k}-{\bf p},\tau
^{\prime })\Phi ({\bf p}^{\prime },\tau ^{\prime \prime })\Phi ({\bf k}%
^{\prime }-{\bf p}^{\prime },\tau ^{\prime \prime })\rangle .
\label{tensorSpectrum_t}
\end{eqnarray}
It's convenient to write the spectrum as a function of $z$:
\begin{eqnarray}
&&\langle h_{ij}(\textbf{k},z)h_{ij}(\textbf{k}',z_{\textbf{k}'})\rangle \nonumber\\
&=&\frac{4}{(2\pi)^3 k^4 M_p^4}\int{dz'\tilde{G}_k(\tau,\tau')}\int{dz''\tilde{G}_k(\tau,\tau'')}  \nonumber\\
&&\times \int{d^3pd^3p'}\Pi^{\ \ kl}_{ij}(\textbf{k})\Pi^{\ \ mn}_{ij}(\textbf{k}')p_k p_l p'_m p'_n  \nonumber\\
&&\times\langle\delta\varphi(\textbf{p},\tau')\delta\varphi(\textbf{k}-\textbf{p},\tau')\delta\varphi(\textbf{p}',\tau'')\delta\varphi(\textbf{k}'-\textbf{p}',\tau'')\rangle,  \nonumber\\
\label{tensorSpectrum_z}
\end{eqnarray}
where the relation $z=k/aH$ have been used and $\tilde{G}$ are defined as
\begin{eqnarray}
    \tilde{G}_k(z,z') = \frac{1}{z'^2}\big[(1-zz')\textrm{sin}(z-z')+(z-z') \nonumber\\ \textrm{cos}(z'-z)\big]\theta(z'-z). \label{GFtilder}
\end{eqnarray}
Now, one should note that $z'_p$ are defined as $z'_p=p/a(\tau')H$. If we omit the subscript like $z''$, it represents a variable in terms of wave number $k$,i.e. $z''=k/a(\tau'')H$.

Using Wick theory, the average of $\langle...\rangle$ becomes
\begin{eqnarray}
&&\langle\delta\varphi(\textbf{p},\tau')\delta\varphi(\textbf{k}-\textbf{p},\tau')
\delta\varphi(\textbf{p}',\tau'')\delta\varphi(\textbf{k}'-\textbf{p}',\tau'')\rangle \nonumber\\
&&=\langle\delta\varphi(\textbf{p},\tau')\delta\varphi(\textbf{k}-\textbf{p},\tau')\rangle
\langle\delta\varphi(\textbf{p}',\tau'')\delta\varphi(\textbf{k}'-\textbf{p}',\tau'')\rangle \nonumber\\
&&+\langle\delta\varphi(\textbf{p},\tau')\delta\varphi(\textbf{p}',\tau'')\rangle
\langle\delta\varphi(\textbf{k}-\textbf{p},\tau')\delta\varphi(\textbf{k}'-\textbf{p}',\tau'')\rangle \nonumber\\
&&+\langle\delta\varphi(\textbf{p},\tau')\delta\varphi(\textbf{k}'-\textbf{p}',\tau'')\rangle
\langle\delta\varphi(\textbf{p}',\tau'')\delta\varphi(\textbf{p}',\tau'')\rangle.  \label{Wick}
\end{eqnarray}
The first term on the right hand can be ignored because it is a diagram containing the disconnected term proportional to $\delta(\textbf{k})\delta(\textbf{k}')$. And it is not hard to find that the second term and the third term are equivalent to each other.

Now the first priority is to solve Eq. \eqref{EOM_pertubed_field_z}. According to \eqref{22} and Appendix \ref{Green_function}, the solution of
\eqref{EOM_pertubed_field_z} is
\begin{eqnarray}
\delta\varphi(\textbf{k},z)=\int{dz'g_k(z,z')\frac{1}{z'^2H^2}\Big[\xi_T(\textbf{k},z')+\xi_q(\textbf{k},z')\Big]}, \nonumber\\ \label{solution_Perturbed_Field_z}
\end{eqnarray}
where
\begin{eqnarray}
g_k(z,z')&=&\frac{z^\nu z'^\nu}{z'^{2\nu}(2/\pi z')}\big[J_\alpha(z)Y_\alpha(z')\nonumber\\
&&\quad\quad\quad-J_\alpha(z')Y_\alpha(z)\big]\theta(z'-z),  \label{GF_Perturbed_Field_z}
\end{eqnarray}
with
\begin{eqnarray}
\nu&=&3(1+Q)/2, \label{nu} \nonumber\\
\alpha&=&\sqrt{\nu^2-\frac{3\beta Q}{1+Q}-3\eta}.
\label{alpha}
\end{eqnarray}
Write perturbed field as
\begin{eqnarray}
\delta\varphi=\delta\varphi_T+\delta\varphi_q,
\end{eqnarray}
where $\delta\varphi_T$ is the part including integral of thermal noise $\xi_T$ while $\delta\varphi_q$ for quantum noise. Then we can find that tensor spectrum contains three \
components: thermal term $\langle\delta\varphi_T\delta\varphi_T\rangle \langle\delta\varphi_T\delta\varphi_T\rangle$, quantum term $\langle\delta\varphi_q\delta\varphi_q\rangle
\langle\delta\varphi_q\delta\varphi_q\rangle$ and cross term $\langle\delta\varphi_q\delta\varphi_q\rangle
\langle\delta\varphi_T\delta\varphi_T\rangle$. Next we will calculate the spectrum in terms of three components above.

\subsection{\label{thermal_term}Thermal term}
Thermal term of tensor spectrum at the end of inflation $(\tau=0,\ z\rightarrow 0)$ reads
\begin{eqnarray}
&&\langle h_{ij}(\textbf{k})h_{ij}(\textbf{k}')\rangle_T \nonumber\\
&=&\frac{4}{(2\pi)^3 k^4 M_p^4}\int{dz'\tilde{G}_k(\tau,\tau')}\int{dz''\tilde{G}_k(\tau,\tau'')}  \nonumber\\
&&\times \int{d^3pd^3p'}\Pi^{\ \ kl}_{ij}(\textbf{k})\Pi^{\ \ mn}_{ij}(\textbf{k}')p_k p_l p'_m p'_n  \nonumber\\
&&\times \langle\delta\varphi_T(\textbf{p},\tau')\delta\varphi_T(\textbf{k}-\textbf{p},\tau')\rangle
\langle\delta\varphi_T(\textbf{p}',\tau'')\delta\varphi_T(\textbf{k}'-\textbf{p}',\tau'')\rangle,  \nonumber\\ \label{thermalTerm}
\end{eqnarray}
The fluctuation-dissipation relation of \eqref{dis_flu_relation_k} acts in terms of $t$. Now using $t=H^{-1}\ln(k/Hz)$ together with
\begin{eqnarray}
\delta(f(x))=\sum_{\{x_0\}}\frac{\delta(x-x_0)}{|f'(x_0)|}, \label{deltaFunction}
\end{eqnarray}
where $x_0$ are zero point of $f(x)$, we obtain
\begin{eqnarray}
\langle \xi_T(\textbf{k},t)\xi_T(\textbf{k}',t')\rangle=2(2\pi^3)\Upsilon T \frac{H^4}{k^2 k'} z^3z' \delta^3(\textbf{k}+\textbf{k}')\delta(z-z'). \nonumber\\
\label{dis_flu_relation_z}
\end{eqnarray}

It is helpful to make such calculation below:
\begin{eqnarray}
&&\langle \delta\varphi_T(\textbf{p},z'_p)\delta\varphi_T(\textbf{p}',z''_{p'})\rangle \nonumber\\
&=&\frac{\pi^2}{4H^2}\int_{z'_p}^\infty{dz1}\int_{z''_{p'}}^\infty{dz_2(z'_{p} z''_{p'})^\nu (z_1z_2)^{-1-\nu}} \nonumber\\
&&\times\tilde{g}_p(z'_{p},z_1)\tilde{g}_{p'}(z''_{p'},z_2)\langle \xi_T(\textbf{p},z_1)\xi_T(\textbf{p}',z_2)\rangle  \nonumber\\
&\simeq & \frac{\pi^5\Upsilon T \delta^3(\textbf{k}+\textbf{k}')}{p^3}\Big[z'^\nu_p Y_\alpha(z'_p)\Big]\Big[z''^\nu_p Y_\alpha(z'_{p'})\Big]  \nonumber\\
&&\times\int_{z'_p}^\infty{dz_1 z^{2-2\nu}_1 J^2_\alpha(z_1)},
\label{briefCalculation}
\end{eqnarray}
where we have used \eqref{Bessel_prop1} and $z^\nu J_\alpha(z)\sim0$ $(z<1)$. Considering the function $z^{2-2\nu}J_\alpha^2(z)$, we find that this function almost equals to zero
except a narrow peak at $z\gtrsim1$, so it is convenient to treat it as delta function. Thus
\begin{eqnarray}
&&\int_{z'_p}^\infty{dz_1 z^{2-2\nu}_1 J^2_\alpha(z_1)} \nonumber\\
&&\simeq \theta(1-\bar{p}z')\int_{0}^\infty{dz_1 z^{2-2\nu}_1 J^2_\alpha(z_1)}  \nonumber\\
&&=\theta(1-\bar{p}z')\frac{\Gamma(\nu-1)\Gamma(\alpha-\nu+3/2)}{2\sqrt{\pi}\Gamma(\nu-1/2)\Gamma(\alpha+\nu-1/2)},   \label{semiCalcu1}
\end{eqnarray}
where $z_p'=p/a(\tau')H=(p/k)(k/a(\tau')H)=\bar{p}z'$ and \eqref{doubleBessel} has been used. With \eqref{nu} and \eqref{alpha}, get $\alpha\approx\nu$. Thus
\begin{eqnarray}
&&\langle\delta\varphi_T(\textbf{p},z'_p)\delta\varphi_T(\textbf{p}',z''_{p'})\rangle \nonumber\\
&=&\frac{\pi^5 \Upsilon T}{p^3}\delta^3(\textbf{p}+\textbf{p}')\big(2^\nu\Gamma(\nu)\big)^2  \nonumber\\
&\times&\frac{\Gamma(\nu-1)\Gamma(\frac{3}{2})}{2\sqrt{\pi}\Gamma(\nu-1/2)\Gamma(2\nu-1/2)}.
\end{eqnarray}

The spectrum damps out as $(\sin k\tau-k\tau\cos k\tau)/k^3$ in Green's functions at large value of $k$. On the other hand, $k$ only appear in the sublimit of the integral and then they are absorbed in $\theta$ function, and $\textbf{k}$ become not so important which act only as the form of $\textbf{k}-\textbf{p}$. Besides,  and the most important, \eqref{coorelation2} indicates that if $p<ak_F$, $\delta\varphi(\textbf{p},t_0)$ has not thermalized during inflation. So the integral in tensor spectrum gets its main contributions at $p\gg k$. With these approximations,
\begin{eqnarray}
&&\langle h_{ij}(\textbf{k})h_{ij}(\textbf{k}')\rangle_T \nonumber\\
&=&\frac{\pi^3 \Upsilon^2 T^2\delta^3(\textbf{k}+\textbf{k}')}{16k^4 M_p^4}\bigg\{\frac{\big[2^\nu\Gamma(\nu)\big]^2\Gamma(\nu-1)}{4\Gamma(\nu-1/2)\Gamma(2\nu-1/2)}\bigg\}^2 \nonumber\\
&&\times  \int_0^\infty{dz'\frac{\sin z'-z'\cos z'}{z'^2}}\int_{z'}^\infty{dz'\frac{\sin z''-z''\cos z''}{z''^2}} \nonumber\\
&&\times \int{d^3pd^3p'\Big(p^2-\frac{(\textbf{p}\cdot\textbf{k})^2}{k^2}\Big)^2}
\frac{\delta^3(\textbf{p}+\textbf{p}')}{p^6}\theta(1-\bar{p}z'). \nonumber \\
    %\label{tensorSpectrum_z}
\end{eqnarray}
Using Eqs. \eqref{projector2}, \eqref{Gamma_prop1}, and \eqref{integral1} $\sim$ \eqref{integral3}, thermal term of tensor spectrum \eqref{thermalTerm} is finally simplified to
%\begin{widetext}
\begin{eqnarray}
&&\langle h_{ij}(\textbf{k})h_{ij}(\textbf{k}')\rangle_T=\frac{3 \pi^4 H^4}{10k^3 M_p^4}\delta^3(\textbf{k}+\textbf{k}')
\bigg(\frac{T}{H}\bigg)^2  \nonumber\\
&&\times\bigg[\frac{Q8^Q\Gamma(3Q/2+3/2)^3}{(3Q+1)\Gamma(3Q/2+1)\Gamma(3Q+5/2)}\bigg]^2. \label{thermalTermFinal}
\end{eqnarray}
%\end{widetext}
Here, we have assumed $z_2>z_1$  and this assumption has no affect on the final result.

\subsection{\label{quantum_term}Quantum term}

According to \eqref{correlationfunctionwarm}, the correlation function of $\delta\varphi_q(\textbf{k},z)$ is
\begin{eqnarray}
&&\langle\delta\varphi_q(\textbf{p},z'_p)\delta\varphi_q(\textbf{p}',z''_{p'})\rangle \nonumber\\
&&=\int_{z'_{p}}^\infty{dz_1}\int_{z''_{p'}}^\infty{dz_2 g(z'_{p},z_1)g(z''_{p'},z_2)} \nonumber\\
&&\times \frac{1}{H^4(z_1z_2)^2}\langle \xi_q(\textbf{p},z_1)\xi_q(\textbf{p}',z_2)\rangle \nonumber\\
&&=\frac{(2\pi)^3}{p^4}[2n(k)+1]\delta^3(\textbf{p}+\textbf{p}')F_p(z_p')F^*_p(z_p''), \label{corelationQuantumTerm}
\end{eqnarray}
where
\begin{eqnarray}
F_p(z_p)=\int_{z_{p}}^\infty{dzG(z_p,z)f_p(z)}.
\end{eqnarray}
Using \eqref{windowfunction}, \eqref{fkz}, \eqref{Bessel_prop1} and \eqref{Neumann_prop1}, $F_p(z_p)$ reads
\begin{eqnarray}
F_p(z_p)\simeq-i\frac{z_p^{3/2-\mu}H}{\sqrt{2}p^3}, \label{Fp}
\end{eqnarray}
with $0<z<\epsilon<1$ \cite{1475-7516-2013-03-032}. Notice that $z_p=p/aH=\bar{p}z<\epsilon$ and $z>\epsilon$, so $0<\bar{p}<1$. Thus the
quantum term of two-point function can be written as
\begin{eqnarray}
\langle h_{ij}(\textbf{k})h_{ij}(\textbf{k}')\rangle_q
\simeq \frac{32\pi^3H^4}{5k^3M_p^4}I\big({H}/{T}\big)\delta(\textbf{k}+\textbf{k}'), \label{correlationQuantermTerm}
\end{eqnarray}
where
\begin{eqnarray}
&&I\left(H/T\right)=\int_0^1{d\bar{p}}\left[\int_0^\infty{}dz\frac{\sin z-z\cos z}{z^2}\right.   \nonumber\\
&&\times \left.\coth\left(\frac{(\bar{p}-\bar{p}_0)zH}{2T}\right)\right]^2. \label{I}
\end{eqnarray}

Next, we will proof that $ I\big(H/T\big)$ is normalized no matter at high temperature or low temperature. Note that $p=(\bar{p}-\bar{p}_0)H/T$ and consider the low temperature condition ($T\ll H$ or $p\gg 1$):
\begin{eqnarray}
&&\int_0^\infty\frac{\sin x-x\cos x}{x^2}\coth\bigg(\frac{px}{2}\bigg) \nonumber\\
&=&\int_0^\infty{dx\frac{\sin x-x\cos x}{x^2}}+\int_0^\infty{\frac{dx}{e^{px}-1}\frac{\sin x-x\cos x}{x^2}} \nonumber\\
&=&1+2\sum_{n=1}^\infty{(-1)^{n-1}\frac{2n}{(2n+1)!}\int_0^\infty{\frac{x^{2n-1}}{e^{px}-1}dx}} \nonumber\\
&=&1-2\sum_{n=1}^\infty{\frac{\zeta(2n)}{2n+1}(ip)^{-2n}} \nonumber\\
&=&1-\ln\Big(\frac{\pi/ip}{\sin(\pi/ip)}\Big)+{ip}\int_{-\frac{1}{ip}}^{\frac{1}{ip}}\ln\Gamma(1+z)dz \nonumber\\
&=&1-\ln\Big(\frac{\pi/p}{\sinh(\pi/p)}\Big)+2\ln\Gamma(1+z)\Big|_{z\rightarrow 0} \nonumber\\
&\simeq & 1.
\label{lowTemperature}
\end{eqnarray}
In the calculations above, we have used relevant properties of Gamma function and Zeta function in Appendix \ref{special_function}. We obtain $I$ is unit at low temperature. Next consider the high temperature condition ($T\gg H$) and define $z=\exp({p}_0/aT)$, $\tilde{p}=\bar{p}H/T$. High temperature means that $p_0$ is a negative number with sufficient large value i.e. $z\ll 1$, thus
\begin{eqnarray}
&&\int_0^\infty\frac{\sin x-x\cos x}{x^2}\coth\bigg(\frac{(\bar{p}-\bar{p}_0)xH}{2T}\bigg) \nonumber\\
&=&\int_0^\infty{dx\frac{\sin x-x\cos x}{x^2}}+\int_0^\infty{\frac{dxze^{-\tilde{p}x}}{1-ze^{-\tilde{p}x}}\frac{\sin x-x\cos x}{x^2}} \nonumber\\
&=&1+2\sum_{n,m=1}^\infty{(-1)^{n-1}\frac{2n}{(2n+1)!}\int_0^\infty{x^{2n-1}\big(ze^{-\tilde{p}x})^m}dx} \nonumber\\
&=&1+2\sum_{n,m=1}^\infty{\frac{(-1)^{n-1}}{2n+1}\frac{1}{\tilde{p}^{2n}}\frac{z^m}{m^{2n}}} \nonumber\\
&=&1+2\sum_{n=1}^\infty{\frac{(-1)^{n-1}}{2n+1}\frac{1}{\tilde{p}^{2n}}g_{2m}(z)} \nonumber\\
&\simeq & 1.
\label{highTemperature}
\end{eqnarray}

Thus we assume that $I(T/H)=1$. This assumption is reliable. In stochastic approach, we assume a field $\phi_q$ as the vacuum fluctuation for
short wavelength with a sharp momentum cutoff, which means  such a fluctuation as a noise always exists no matter at low temperature or high
temperature. It is just the condition for cold inflation. On the other hand, chemical potential is a physical variable that cannot be ignored
when a particles coupling with other fields (like thermal bath) especially for the condition at phase transition. Most importantly, it is just
the chemical potential that eliminates the singularity at $p=0$.

\subsection{\label{cross_term}Cross term}

If we simulate the calculation above, it is not hard to get the expression of the cross term for the two-point correlation function:
\begin{eqnarray}
&&\langle h_{ij}(\textbf{k})h_{ij}(\textbf{k}')\rangle_{\textrm{cross}}
\simeq\frac{3\pi^4H^4}{5k^3M_p^4}\delta(\textbf{k}+\textbf{k}') \nonumber\\ &&\times\frac{T}{H}\frac{Q8^Q\Gamma(3Q/2+3/2)^3}{(3Q+1)\Gamma(3Q/2+1)\Gamma(3Q+5/2)}. \label{correlationCrossTerm}
\end{eqnarray}
This component has nothing new compared with thermal and quantum component.

In the calculations above, we have ignored the slow parameter as the index of variable $z$. With this approximation, we obtain the scale-invariant spectrum of each components.

\section{\label{result}Result and Numerical Analysis}

\begin{figure}
  \centering
  \includegraphics[width=3.6in,height=3in]{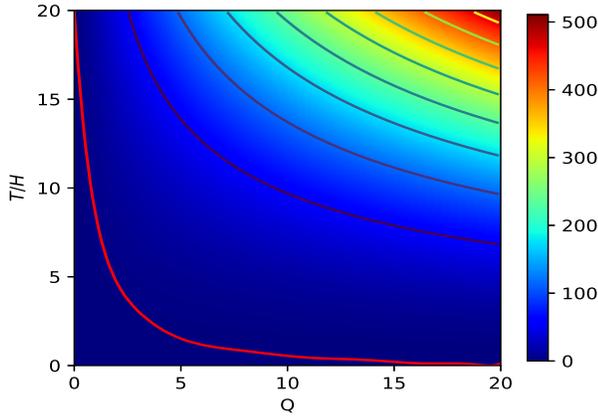}
  \caption{We plot the thermal component of tensor perturbation which is normalized as $H^4/M_p^4$. The spectrum increases with the increase of $\Upsilon/H$ and $T/H$. The bright red line on the lower left quarter is the contour line with the value of amplitude of quantum fluctuation for tensor mode. The amplitude of thermal component below the contour line is smaller than the value on the contour line£¬i.e.£¬$P_{h,T}/P_{h,q}<1$. In other words, quantum fluctuations dominate at this epoch. The condition is opposite above the contour line, which is almost the same with the work of Ramos\cite{1475-7516-2013-03-032}.}\label{thermal}
\end{figure}

\begin{figure}
  \centering
  \includegraphics[width=3.6in,height=3in]{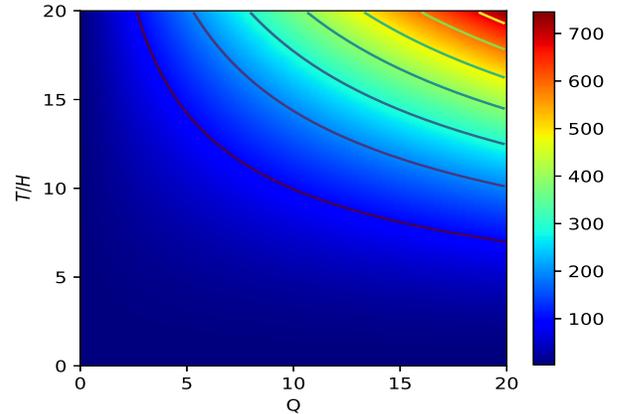}
  \caption{Total spectrum for tensor perturbation.}\label{total}
\end{figure}

Power spectrum of tensor perturbation $P_h(\textbf{k})$ is defined as:
\begin{eqnarray}
\langle h_ij(\textbf{k})h_ij(\textbf{k}')\rangle=\frac{2\pi^2}{k^3}P_h(\textbf{k})\delta(\textbf{k}+\textbf{k}'). \label{PS}
\end{eqnarray}
The spectrum of tensor modes with vacuum form reads \cite{2003moco.book.....D}
\begin{eqnarray}
P_{h,\textrm{vac}}(\textbf{k})=\frac{2}{\pi^2}\frac{H^2}{M_p^2}. \label{vacummTensor}
\end{eqnarray}
We illustrate the power spectrum for thermal component in Fig.~\ref{thermal} and total spectrum of tensor perturbation for warm inflation in
Fig.~\ref{total}.

As discussed above, the spectrum for quantum component is  a constant with order of $H^4/M_p^4$. This value is so small that there is almost no correction to the spectrum of primordial tensors, which agrees with other cold inflation models \cite{PhysRevD.75.123518,PhysRevD.85.023534, PhysRevD.88.103518}. However, what we interest most is the thermal component in total spectrum. As is shown in \eqref{thermalTermFinal}, there are two variables determining the thermal spectrum i.e. $\Upsilon/H$ and $T/H$. Considering the weak dissipation condition with $\Upsilon/H\ll 1$ and $T/H\ll 1$ first, we can directly get $P_{h,T}(\textbf{k})\rightarrow 0$. In other words, this condition is just the same with cold inflation. In contrast, the condition with strong dissipation ($\Upsilon>H$ and $T>H$) is quite different from that of weak dissipation. The amplitude of thermal spectrum increases with the increase of $\Upsilon/H$ and $T/H$. Taking two simple examples, $P_{h,T}(\textbf{k})=63.87$ at $\Upsilon/H=10$ and $T/H=10$, or especially, $P_{h,T}(\textbf{k})\sim 10^4$ at
$\Upsilon/H=100$ and $T/H=100$. It has been widely studied that there has a significant correction on the spectrum of primordial
fluctuations in the warm regime with strong dissipation for $T>H$ and such a correction lowers the tensor-to-scalar ratio. Thus the curvature
power spectrum is modified into \cite{PhysRevLett.117.151301,2014PhLB..732..116B}
\begin{eqnarray}
     \Delta^2_\textrm{R}=\Big(\frac{H}{\dot{\phi}}\Big)^2\Big(\frac{H}{2\pi}\Big)^2\Big[1+2n+2\pi Q\frac{T}{H}\Big]. \label{curvature}
\end{eqnarray}
The temperature at the end of inflation is \cite{2016JCAP...11..022G}
\begin{eqnarray}
4.09\times 10^{13}\textrm{GeV}\leq T\leq 2.216\times 10^{14}\textrm{GeV}. \label{temperature}
\end{eqnarray}
With this and using \eqref{thermalTermFinal}, \eqref{vacummTensor} and \eqref{curvature}, we obtain the tensor-to-scalar ratio in warm inflation:
\begin{eqnarray}
r\simeq\frac{[10^{-8}f(Q)+8]|n_T|}{1+2n+2\pi Q\frac{T}{H}}, \label{ratio}
\end{eqnarray}
where $f(Q)$ is function containing $Q$ in \eqref{thermalTermFinal} and $n_T$ is the tensor index. From \eqref{ratio} we can find that large $Q$ will upper the tensor-to-scalar ratio , but it does not act obviously until $Q$ reaches the level of 100 ($Q>100$). In
\cite{2017PhRvD..95b3517B}, the authors have tabled the parameters like $Q$, $T/H$, $r$ and others in warm inflation with different potential
by the constraint by recent years' observational data. The result shows $T/H\sim1$ during inflation which agrees with the analysis in
\cite{2016JCAP...11..022G}. The nondetection of cosmic gravitational waves background strongly constrains that thermal component of tensor
perturbation $P_{h,\textrm{T}}$ will not upper the ratio dramatically which also agrees with a various of works
\cite{2017PhRvD..95b3517B,PAGANO2016823}. In this way, the primordial tensor spectrum thus can be used to distinguish warm inflation from
cold inflation.

\section{\label{conclusion}CONCLUSIONS AND DISCUSSION}

In this paper we first derive the two point correlation function of tensor perturbation in warm inflation and prove that primordial scale-invariant power spectrum of warm inflation is achieved only if system evolutes near thermal equilibrium state, in which condition warm inflaton fields are stable. We also explain the physical meaning of both scalar index and slow-roll parameter $\beta$ in \eqref{beta} that correspond to the nonequilibrium properties during inflationary epoch. Then we calculate the power spectrum for warm inflation by Green's function method and mainly discuss the thermal component in total spectrum. We consider it as a new method to distinguish warm inflation from cold inflation. At last, we illustrate our result by numerical analysis. Using existing cosmic observational data, we find that fluctuations from thermal noise does not raise the tensor-to-scalar ratio dramatically at order of $10^{-4}$ although the temperature is high enough.

There are also many issues which deserve further discussion in this paper. The first one is the thermal properties of warm inflation model. For example, how initial condition determines the scalar index $n_s$ and whether the stability of thermal inflaton fields still holds with arbitrary initial condition? Then, we find that the spectrum of quantum fluctuation is almost a constant because of momentum cutoff in stochastic approach. So whether quantum fluctuations still exist in high temperature in other models becomes an interesting and challenging work. Finally, but the most important, is whether we can observe such a fluctuation? Although the cosmic gravitational waves background has not been detected yet, with the
discovery of gravitational waves, more and more new method coming out \cite{PhysRevLett.118.091102, 1475-7516-2015-12-006} and a series of
observation installations working or being build \cite{PhysRevX.6.041015, VIRGO, GEO600, TAMA300}, it is believable that primordial
gravitational waves can be detected in the foreseeable future.

\begin{acknowledgments}
This work was supported by the National Natural Science Foundation of China (Grants No. 11575270,
No. 11175019, and No. 11235003). The authors would like to thank Zong-Kuan Guo and Zong-Hong Zhu for helpful comments and discussions.
\end{acknowledgments}

\appendix

\section{\label{spectrum}Thermal properties of warm inflation}

Correlation function is defined as
\begin{eqnarray}
\mathcal{P}_{\delta\varphi}(\textbf{x}-\textbf{y},t_1,t_2)=\langle\delta\varphi(\textbf{x},t_1)\delta\varphi(\textbf{y},t_2)\rangle,
\label{correlationFunction}
\end{eqnarray}
whose Fourier transformation is
\begin{eqnarray}
\mathcal{P}_{\delta\varphi}(\textbf{k},t_1,t_2)=\int{}\frac{d^3k'}{(2\pi)^3}\langle\delta\varphi(\textbf{k},t_1)
\delta\varphi(\textbf{k}',t_2)\rangle,
\label{correlationFunctionk}
\end{eqnarray}

Let's consider (\ref{EOM_pertubed_field}) again. Using the slow-roll condition and strong dissipation condition $\Gamma\gg H$, \eqref{EOM_pertubed_field} approximately writes  \cite{2000NuPhB.585..666B}
\begin{eqnarray}
\Upsilon\frac{d\delta\varphi(\textbf{k},t)}{dt}+[k^2_p+V''(\phi)]\delta\varphi(\textbf{k},t) \approx \xi_T(\textbf{k},t), \label{approx_EOM}
\end{eqnarray}
where $V''(\phi)=d^2V(\phi)/d\phi^2$ and $\phi$ is defined in ~(\ref{BGinflaton}), $k_p=k/a$ is the physical wave number and $k$ is the conformal wave number. Strong dissipation means that we can ignore the change of parameter $a$, $k_p$ and $T$ within the time interval $1/H$ and fluctuation from quantum noise is negligible. Then the solution of ~(\ref{EOM_pertubed_field}) is
\begin{eqnarray}
\delta\varphi(\textbf{k},t)&\approx &\frac{1}{\Upsilon}e^{-(t-t_0)/\tau(\phi)}\int^t_{t_0}{e^{t'-t_0}\xi_t(\textbf{k},t')dt'} \nonumber\\
&&+\delta\varphi(\textbf{k},t_0)e^{-(t-t_0)/\tau(\phi)} \label{approx_solution}
\end{eqnarray}
where $t_0$ is any coordinate time during inflation and $\tau(\phi)=\Upsilon/[k^2_p+V''(\phi)]$.

The correlation function of perturbed inflation field is
\begin{widetext}
\begin{eqnarray}
\langle\delta\varphi(\textbf{k},t_1)\delta\varphi(\textbf{k}',t_2)\rangle &=&\delta\varphi(\textbf{k},t_0)\delta\varphi(\textbf{k}',t_0)
e^{-(t_1+t_2)/\tau(\phi)}+\frac{2(2\pi)^3T\delta^3(\textbf{k}+\textbf{k}')}{\Upsilon a^3}e^{-(t_1+t_2)/\tau(\phi)}
\nonumber\\
&&\quad \times\int^{t_1}_{t_0}\int^{t_2}_{t_0}{e^{(s_1+s_2)/\tau(\phi)}\delta(s_1-s_2)ds_1 ds_2} \label{correlation1}
\end{eqnarray}
where we have used ~(\ref{dis_flu_relation_k}). The double integral in ~(\ref{correlation1}) contains a $\delta$ function, so we need to integrate first to the lager one in $t_1$ and $t_2$. Then
\begin{eqnarray}
\langle\delta\varphi(\textbf{k},t_1)\delta\varphi(\textbf{k}',t_2)\rangle &=& \delta\varphi(\textbf{k},t_0)\delta\varphi(\textbf{k}',t_0)
e^{-(t_1+t_2)/\tau(\phi)}+\frac{2(2\pi)^3T\delta^3(\textbf{k}+\textbf{k}')}{\Upsilon a^3}e^{-(t_1+t_2)/\tau(\phi)}  \nonumber\\
&& \times\int^{\textrm{min}(t_1,t_2)}_{t_0}\int^{\textrm{max}(t_1,t_2)}_{t_0}{e^{(s_1+s_2)/\tau(\phi)}\delta(s_1-s_2)ds_1 ds_2} \nonumber\\
&=&\delta\varphi(\textbf{k},t_0)\delta\varphi(\textbf{k}',t_0)e^{-(t_1+t_2)/\tau(\phi)}+\frac{2(2\pi)^3T\delta^3(\textbf{k}+\textbf{k}')}
{\Upsilon a^3}e^{-(t_1+t_2)/\tau(\phi)}\int^{\textrm{min}(t_1,t_2)}_{t_0}{e^{{2s}/{\tau(\phi)}}ds} \nonumber\\
&=& \delta\varphi(\textbf{k},t_0)\delta\varphi(\textbf{k}',t_0)e^{-(t_1+t_2)/\tau(\phi)}+
\frac{(2\pi)^3T\tau(\phi)\delta(\textbf{k}+\textbf{k}')}
{\Upsilon a^3}\Big\{e^{-[t_1+t_2-2\textrm{min}(t_1,t_2)]/\tau(\phi)}-e^{-(t_1+t_2)/\tau(\phi)}\Big\}  \nonumber\\
&=& \delta\varphi(\textbf{k},t_0)\delta\varphi(\textbf{k}',t_0)e^{-(t_1+t_2)/\tau(\phi)}+
\frac{(2\pi)^3T\tau(\phi)\delta^3(\textbf{k}+\textbf{k}')}{\Upsilon a^3}\Big\{e^{-|t_1-t_2|/\tau(\phi)}-e^{-(t_1+t_2)/\tau(\phi)}\Big\}. \label{coorelation2}
\end{eqnarray}

Now, let's make more detailed calculation of the second term in final equation.
\begin{eqnarray}
\frac{(2\pi)^3T\tau(\phi)}{\Upsilon a^3}
=\frac{(2\pi)^3T\Upsilon}{a^3\Upsilon(k^2/a^2+V'')}  
=\frac{(2\pi)^3HT}{a^3H^3(\frac{k^2}{a^2H^2}+\frac{V''}{H^2})}
= \frac{(2\pi)^3 HT}{\frac{k^3}{z^3}(z^2+3\eta)}
\simeq \frac{(2\pi)^3 HTz}{k^3}, \label{correlation3}
\end{eqnarray}
where we have used relation $z=k/aH$ and Eq.~(\ref{eta}). Finally, the correlation function reads
\begin{eqnarray}
\langle\delta\varphi(\textbf{k},t_1)\delta\varphi(\textbf{k}',t_2)\rangle=\Big(\delta\varphi(\textbf{k},t_0)\delta\varphi(\textbf{k}',t_0)
-\frac{(2\pi)^3 HTz}{k^3}\delta^3(\textbf{k}+\textbf{k}')\Big)e^{-(t_1+t_2)/\tau(\phi)}+\frac{(2\pi)^3 HTz}{k^3}\delta^3(\textbf{k}+\textbf{k}')e^{-|t_1-t_2|/\tau(\phi)}.  \nonumber\\
\label{correlation4}
\end{eqnarray}
\end{widetext}

Thermalization requires that initial condition must thermalize at physical scale $k_p$ within $1/H$. From ~(\ref{correlation4}), the first term in right hand must be negligible with in Hubble time, so
\begin{eqnarray}
\frac{k^2_p-V''}{H\Upsilon}>1. \label{thermalization_condition}
\end{eqnarray}
If $V''>H\Upsilon$, the freeze-out number is the same with that of cold inflation, $k_F=H$ ($z=1$). However, if $V''<H\Upsilon$, the freeze-out number reads
\begin{eqnarray}
k_F=(H\Upsilon)^{\frac{1}{2}}. \label{freeze_out_number}
\end{eqnarray}

Obviously $\mathcal{P}(\textbf{k},t_1,t_2)$ is dependent on the initial state $\delta\varphi(\textbf{k},t_0)$. Now make average on initial state
and assume $t_1=t_2=t$ together with $t_0=0$, so correlation function can be written as
\begin{eqnarray}
\mathcal{P}_{\delta\varphi}(\textbf{k}',t_0)=\Big(\frac{2\pi^2}{k^3}P_{\delta\varphi}(\textbf{k}',t_0)-\frac{HTz_*}{k^3}\Big)
e^{-2t/\tau{\phi}}
+\frac{HTz_*}{k^3}, \nonumber\\ \label{correlationFinal}
\end{eqnarray}
with the definition of power spectrum
\begin{eqnarray}
P_{\delta\varphi}(\textbf{k},t)=\frac{k^3}{2\pi^2}\int{\frac{d^3k'}{(2\pi)^3}\langle\delta\varphi(\textbf{k},t_0)
\delta\varphi(\textbf{k}',t_0)\rangle}, %\label{powerspectrum}
\end{eqnarray}
and $z_*$ as the value at horizon crossing $z_*=(\Upsilon H)^{1/2}/H$.
Things become quite interesting. If the power spectrum is scale-invariant i.e.  $\mathcal{P}_{\delta\varphi}(\textbf{k},t_0)=(\Upsilon H)^{1/2}
T/2\pi^2$ \cite{PhysRevD.69.083525}, the spectrum $P_{\delta\varphi}(\textbf{k},t)$ is also scale-invariant and totally the same with
$\mathcal{P}_{\delta\varphi}(\textbf{k},t_0)$. In other words, $P_{\delta\varphi}$ is stable in this condition which means system is on thermal
equilibrium during time interval $t>t_0$. This is quite similar with the condition of correlation function for particles with Brown motion
\cite{2006stme.book.....S}. Now, consider that $P_{\delta\varphi}(\textbf{k},t_0)$ is not a scale-invariant spectrum and assume
$P_{\delta\varphi}(\textbf{k},t_0)=A(k/k_0)^{n'-1}$ where $n'$ is an arbitrary number. Then $P_{\delta\varphi}(\textbf{k},t)$becomes also
dependent on $k$ i.e. $P_{\delta\varphi}(\textbf{k},t)=A(k/k_0)^{n-1}$. However, this term damps with the increase of time, which means $n$
tends to unit with the time evolution. This is an effect dominated by nonequilibrium mechanics. In this way, we can say that scalar index $n_s$
and slow-roll parameter $\beta$ are parameters that illustrate the deviation from equilibrium state. Probe on Cosmic Microwave Background shows
that our universe is almost in nearly equilibrium \cite{WMAP} if considering our universe in early epoch as a model in thermal bath. The
relation \eqref{correlationFinal} indicates that $P_{\delta\varphi}(\textbf{k},t_0)$ (the initial condition of universe) becomes not so
important even though we cannot give an accurate description till now.

\section{\label{Green_function}A brief introduction to Green's Function }

Let's consider the equation \cite{MathemicalMethed}
\begin{subequations}\label{equation}
\begin{eqnarray}
\frac{d^2 y}{d t^2}+\omega^2(t)y=f(t),
\end{eqnarray}
with boundary condition
\begin{eqnarray}
y(0)=y'(0)=0. \label{boundaryofequation}
\end{eqnarray}
\end{subequations}
To solve this equation, we need to solve another relevant equation in terms of Green function G(t,t'):
\begin{subequations}\label{Greenfunction}
\begin{eqnarray}
\frac{d^2 G(t,t')}{d t^2}+\omega^2(t)G(t,t')=\delta(t-t'),
\end{eqnarray}
with boundary condition
\begin{eqnarray}
G|_{t=0}=\frac{d G(t,t')}{dt}\Big|_{x=0}=0. \label{boundaryofGF}
\end{eqnarray}
\end{subequations}
Obviously, when $t<t'$, $G$ satisfy
\begin{eqnarray}
\frac{d^2 G(t,t')}{d t^2}+\omega^2(t)G(t,t')=0, \label{linearGF}
\end{eqnarray}
with boundary condition \eqref{boundaryofGF}. Then assume the general solution of \eqref{linearGF} is
\begin{eqnarray}
G(t<t')=c_1(t')y_1(t)+c_2(t')y_2(t), \label{generalsolutionoflinearGF}
\end{eqnarray}
where $y_1$ and $y_2$ is the solutions of \eqref{linearGF}. Considering the boundary condition, we obtain
\begin{eqnarray}
G(t<t')=0. \label{solutionoflinearGF}
\end{eqnarray}
Next, assume
\begin{eqnarray}
G(t>t')=c_3(t')y_1(t)+c_4(t')y_2(t). \label{generalSolutionNonlinearGF}
\end{eqnarray}
It's not hard to find that $G$ is continuous at $t=t'$ while $G'$ not, and
\begin{eqnarray}
\frac{d G(t,t')}{d t}\Big|_{t<t'}=0,\quad \frac{d G(t,t')}{d t}\Big|^{t+0}_{t-0}=1.
\end{eqnarray}
Thus
\begin{eqnarray}
c_3(t')y_1(t')+c_4(t')y_2(t')=0 \\
c_3(t')y'_1(t')+c_4(t')y'_2(t')=1
\end{eqnarray}
Then we obtain the solution of \eqref{Greenfunction}
\begin{eqnarray}
G(t,t')=\frac{y_1(t)y_2(t')-y_2(t)y_1(t')}{y_1(t)y'_2(t)-y_2(t)y'_1(t)}\theta(t-t'). \label{solutionGF}
\end{eqnarray}
Finally, the solution of \eqref{equation} is
\begin{eqnarray}
y(t)=\int_0^\infty{G(t,t')f(t')dt'}. \label{solutionEquation}
\end{eqnarray}

\section{\label{special_function}Special functions and integrals}
The equation
\begin{eqnarray}
\frac{d^2y}{dz^2}+\frac{1}{z}\frac{dy}{dz}+(1-\frac{\nu^2}{z^2})y=0 \label{equationofBessel}
\end{eqnarray}
has two linear independent solutions: Bessel function $J_\nu(z)$ and Neumann function $Y_\nu(z)$, with properties
\cite{1970hmfw.book.....A,SpecialFunctions}
\begin{eqnarray}
J_\nu(z) \approx \frac{1}{\Gamma(\nu+1)}\Big(\frac{z}{2}\Big)^\nu\ (\nu>0,\ z\rightarrow 0^+), \label{Bessel_prop1}
\end{eqnarray}
\begin{eqnarray}
Y_\nu(z) \approx -\frac{\Gamma(\nu)}{\pi}\Big(\frac{2}{z}\Big)^\nu\ (\nu>0,\ z\rightarrow 0^+), \label{Neumann_prop1}
\end{eqnarray}
and
\begin{eqnarray}
J_\nu(z)Y'_\nu(z)-J'_\nu(z)Y_\nu(z)=\frac{2}{\pi z}. \label{Bessel_prop2}
\end{eqnarray}
The solution of equation
\begin{subequations}\label{22}
\begin{eqnarray}
\frac{d^2u}{dz^2}+\frac{1-2\alpha}{z}\frac{du}{dz}+\Big(\beta^2+\frac{\alpha^2-\nu^2}{z^2}\Big)u = 0 \label{equationof22}
\end{eqnarray}
is also in terms of Bessel function
\begin{eqnarray}
u=z^\alpha Z_\nu(\beta z), \label{solutionof22}
\end{eqnarray}
\end{subequations}
where $Z_\nu$ is any kind of Bessel function. Integral with double Bessel function is Schafgeitlin integral formula:
\begin{eqnarray}
\int_0^\infty& &{\frac{J_\mu(ax)J_\nu(bx)}{x^\lambda}dx}=\frac{b^{\nu}\Gamma(\frac{\nu+\mu-\lambda+1}{2})}{2^\lambda a^{\nu-\lambda+1}
\Gamma(\nu+1)\Gamma(\frac{\mu-\nu+\lambda+1}{2})}\nonumber \\
& &\times F\Big(\frac{\nu+\mu-\lambda+1}{2},\frac{\nu-\mu-\lambda+1}{2};\nu+1;\frac{b^2}{a^2} \Big),\nonumber \\
& &\qquad \mu+\nu+1>\lambda>-1,0<b<a\quad \textrm{or}  \nonumber \\ & &  \quad\quad \mu+\nu+1>\lambda>0, a=b .\label{Schafgeitlin}
\end{eqnarray}
Specially,
\begin{eqnarray}
& & \int_{0}^\infty{dz_1 z^{2-2\nu}_1 J^2_\alpha(z_1)} \nonumber \\
& & = \frac{(1/2)^{2\nu-3}\Gamma(2\nu-2)\Gamma(\alpha-\nu-3/2)}{2[\Gamma(\nu-1/2)]^2\Gamma(\alpha+\nu-1/2)} \nonumber \\
& & = \frac{\Gamma(\nu-1)\Gamma(\alpha-\nu+3/2)}{2\sqrt{\pi}\Gamma(\nu-1/2)\Gamma(\alpha+\nu-1/2)}, \label{doubleBessel}
\end{eqnarray}
where we have used \eqref{Gamma_prop2}. Zeta function is defined as
\begin{eqnarray}
\zeta(z)=\frac{1}{\Gamma(z)}\int_0^\infty{\frac{x^{z-1}}{e^x-1}dx}, \label{zeta}
\end{eqnarray}
where $\Gamma(z)$ are Gamma functions with
\begin{eqnarray}
\Gamma(z+1)=& &z\Gamma(z)\ (\textrm{Re}(z)>1), \label{Gamma_prop1} \\
\Gamma(2z)=2^{2z-1}& &\pi^{-1/2}\Gamma(z)\Gamma(z+{1}/{2}). \label{Gamma_prop2}
\end{eqnarray}
Zeta function has the property \cite{SRIVASTAVA1988129}
\begin{eqnarray}
\sum_{k=1}^\infty{\zeta(2k)}\frac{z^{2k}}{2k+1}&=&\frac{1}{2}\ln\Big(\frac{\pi z}{\sin \pi z}\Big) \nonumber\\ 
&&-\frac{1}{2z}\int_{-z}^z{}\ln\Gamma(1+z)dz,\quad |z|<1. \nonumber\\
\label{zeta_prop}
\end{eqnarray}
We have used the two integral formulas below frequently in this paper
\begin{eqnarray}
\int_0^\infty{dx\frac{\sin x-x\cos x}{x^2}}=1,\label{integral1} \\
\int_0^\infty{dx\frac{\sin x-x\cos x}{x^3}}=\frac{\pi}{4}£¬\label{integral2}
\end{eqnarray}
and
\begin{eqnarray}
\int_0^\infty{\frac{\sin x-x\cos x}{x^3}}\int_0^x{\frac{\sin y-y\cos y}{y^2}dydx}=\frac{\pi}{12} \nonumber\\ \label{integral3}
\end{eqnarray}

%\bibliography{GW}

\end{document}